\documentclass{JHEP}
\input epsf
\usepackage{epsfig}
\usepackage{amssymb}
\usepackage[active]{srcltx}

\newcommand{\bea}{\begin{eqnarray}}
\newcommand{\eea}{\end{eqnarray}}
\newcommand{\bean}{\begin{eqnarray*}}
\newcommand{\eean}{\end{eqnarray*}}

\def\O #1{\overline{#1}}

\def\W #1{\widetilde{#1}}

\def\braket#1{\left\langle #1 \right\rangle}

\def\ket#1{\left| #1\right\rangle}
\def\gb #1{ \left\langle #1 \right]}
\def\tgb #1{ \left[ #1 \right\rangle}
\def\vev#1{\left\langle #1 \right\rangle}
\def\cb #1{ \left[ #1 \right]}

\def\a{{\alpha}}

\def\la{\lambda}
\def\eps{\epsilon}

\def\Label#1{\label{#1}}

\preprint{ITFA-2007-53}
\title{Integral Coefficients for One-Loop Amplitudes}
\author{Ruth Britto$^{a}$ and Bo Feng$^{b}$\\
~~~~\\
$^a$Institute for Theoretical Physics, University of Amsterdam \\
Valckenierstraat 65, 1018 XE Amsterdam, The Netherlands\\
$^b$Center of Mathematical Science, Zhejiang University, Hangzhou, China\\
}
\abstract{ We present a set of algebraic functions for evaluating
the coefficients of the scalar integral basis of a general one-loop
amplitude. The functions are derived from unitarity cuts, but the
complete cut-integral procedure has been carried out in generality so
that it never needs to be repeated.
Where the master integrals are known explicitly, the results
here can be used as a black box with tree-level amplitudes as input
and one-loop amplitudes as output.
}

\begin{document}

\section{Introduction}

Observations of new physics at the Large Hadron Collider will come
from the analysis of many scattering processes with complex final states.
It is
essential to prepare the theoretical groundwork by performing
precision calculations of Standard Model production processes
incorporating at least
 next-to-leading order QCD corrections.  This effort requires efficient
algorithms for computing multi-leg one-loop amplitudes.

The unitarity method introduced in~\cite{Bern:1994zx} is designed to
compute amplitudes by applying a unitarity cut to an amplitude on
one hand, and its expansion in a basis of master integrals on the
other \cite{'t Hooft:1978xw,Bern:1992em,Bern:1993kr,Tarasov:1996br}.  From knowledge of the basis and the general
structure of the coefficients in the expansion, the coefficients can
be constrained.

The holomorphic anomaly \cite{Cachazo:2004by}
reduces the problem of phase space integration to one of algebraic
manipulation, namely evaluating residues of a complex function.  By
applying this operation within the unitarity method, coefficients can
be extracted systematically.  The reason this is possible is that the
unitarity cuts of master integrals  are uniquely identifiable as
analytic expressions.
Accordingly, a method was introduced to evaluate any finite
four-dimensional unitarity cut and systematically derive compact
expressions for the coefficients~\cite{Britto:2005ha,Britto:2006sj}.
The evaluation was carried out in the context of the spinor formalism \cite{Berends:1981rb,De
  Causmaecker:1981bg,Kleiss:1985yh,Gastmans:1990xh,Xu:1986xb,Gunion:1985vca}.
In \cite{Britto:2006fc}, we wrote down these general, compact formulas for
master integral coefficients.

The main purpose of the present paper is to improve upon those formulas in two respects.
First, the coefficients were written as residues of the explicit
formulas in \cite{Britto:2006fc}.  Identifying the residue of a
function simply involves performing a series expansion, but within the spinor
formalism, this expansion is not very transparent in the case
of multiple poles.  Additional instructions were given to aid in
automatizing this step.
Here, our formulas will be given in terms of a truncated series
expansion in a single scalar variable.
Second, the starting point of the formulas  of
\cite{Britto:2006fc}, from which to take input data, was the result of
some spinorial manipulations of the initial cut integrand.  Here, the
input data are determined directly from the initial cut integrand, as
assembled from tree-level amplitudes.

In this paper, we have thus eliminated the need for applying any analytic
spinor identities.  Programming the final formulas is completely
straightforward.  The values of the coefficients are of course
identical to those from the formulas of our previous paper.
Other general expressions for coefficients derived from unitarity cuts and
generalized unitarity cuts of one-loop amplitudes have
been given in \cite{Britto:2004nc,Bern:2004ky,Mastrolia:2006ki,Forde:2007mi,BjerrumBohr:2007vu}.

Starting from analytic expressions for color-ordered tree amplitudes,
we set up the unitarity cut integral.  If $K$ is the momentum in the
unitarity cut, then the two cut propagators can be denoted by $p$ and
$p-K$.  See Figure \ref{fig:cut}.
 \EPSFIGURE[htb]{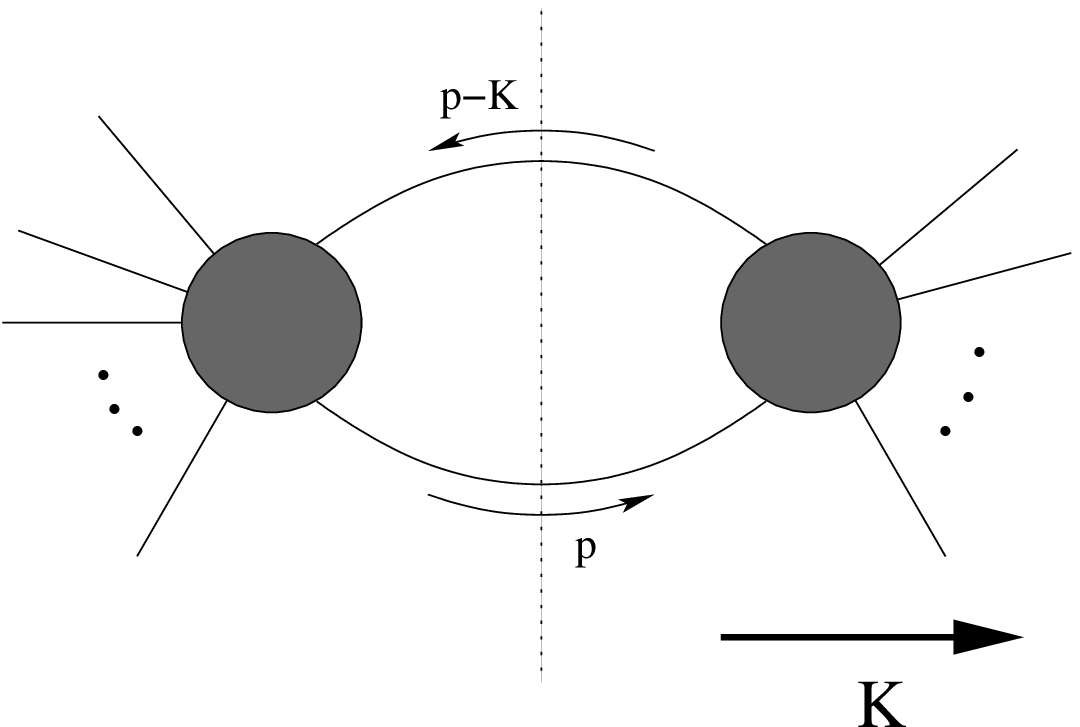}{Representation of the cut integral.  $K$ is
   the sum of external momenta on one side of the cut.  \label{fig:cut}}
In terms of its dependence on the loop momentum $p$, the cut integral is a sum of terms of the following form:
\bea  C & = & c \int d^{4-2\eps} p~ \frac{ \prod_{i=1}^{m} (-2 p
\cdot P_i ) }{\prod_{j=1}^k (p-K_j)^2} \delta^{(+)}(p^2)
\delta^{(+)}((p-K)^2)~~~\Label{I-inte}\eea
Here, $c$ is the prefactor independent of $p$ (but may depend on
$\mu^2$ (or $u$ discussed below) in our dimension regularization) ,
and the values of the momentum vectors $K_j$ are sums of momenta of
cyclically adjacent external particles. We work in the
four-dimensional helicity scheme, so that all external momenta $K_i$
are $4$-dimensional and only the internal momentum $p$ is
$(4-2\eps)$-dimensional.
 Hence we decompose the loop momentum as \cite{Bern:1995ix,Bern:1995db}
\bea
p=\W \ell+\vec{\mu},~~~~\Label{fdh}
\eea
where $\W \ell$ is $4$-dimensional and $\vec{\mu}$ is
$(-2\eps)$-dimensional, and we further define the
extra-dimensional parameter $u$ by
\bea
 u= {4\mu^2\over K^2}.~~~~\Label{def-u}
\eea
Let us then define the following four-vectors:
\bea Q_j & = &  - (\sqrt{1-u})K_j+ \frac{K_j^2- (1-\sqrt{1-u})( K_j\cdot K)}{K^2} K, \nonumber \\
R_i & = & - (\sqrt{1-u}) P_i- \frac{ (1-\sqrt{1-u}) ( P_i \cdot K)}{ K^2}
K.~~~~\Label{RQ} \eea
In terms of these momentum vectors, the cut integral may be
expressed as
\bea C & = & c \int_0^1 du~ u^{-1-\eps}  \int
\vev{\ell~d\ell}[\ell~d\ell] (\sqrt{1-u}) \frac{(K^2)^{n+1}}{
\gb{\ell|K|\ell}^{n+2}} \frac{ \prod_{i=1}^{n+k}
\gb{\ell|R_i|\ell}}{\prod_{j=1}^k
\gb{\ell|Q_j|\ell}}~~~~\Label{cut-refined}\eea
where we have set $n=m-k$, and $\ket{\ell}$ and $|\ell]$ are homogeneous spinors. This follows from the basic steps of
spinor integration, which are reviewed in Appendix
\ref{spinorintegration} but are not needed to apply our formulas for
coefficients. The point is that the only thing  we need to do is
treat a general integrand of the form
\bea (\sqrt{1-u}) (K^2)^{1+n}{1\over \gb{\ell|K|\ell}^{n+2}}
{\prod_{j=1}^{k+n} \gb{\ell|R_j |\ell} \over \prod_{i=1}^k
\gb{\ell|Q_i |\ell}}~.~~~\Label{I-type}\eea
The result of this integration is the subject of this paper.  In
terms of the vectors defined in (\ref{RQ}) from the initial data of
(\ref{I-inte}), and the two integers $k$ and $n$, we give formulas
for the four-dimensional coefficients. For renormalizable theories,
$n \leq 2.$  Terms with $n \leq -2$ contribute to box integrals
only; terms with $n = -1$ contribute to triangle and box integrals;
and terms with $n \geq 0$ contribute to bubble, triangle and  box
integrals. To proceed to the full $d$-dimensional coefficients,
including those for pentagons, one would perform the final integral over $u$
with the recursion and reduction formulas of
\cite{Anastasiou:2006jv, Anastasiou:2006gt}.

We wish to remark on a few features of our formulas.
\begin{itemize}

\item Our starting point is the most general expression in field theory
with a unitarity cut of a one-loop amplitude. Particles can be
massless or massive, although in this paper we focus on massless
propagators. Generalization to massive propagators should be
straightforward.  The propagator can be a scalar, fermion, or
vector, as long as the proper degrees of freedom are accounted
for.\footnote{One way to do this is to use Feynman diagrams
to write down the full one-loop integrand expression, then multiply by
$(p^2-m_1^2)^{-1}((p-K)^2-m_2^2)^{-1}$ along with the  two delta-functions
$\delta(p^2-m_1^2)\delta((p-K)^2-m_2^2)$, and use these to set e.g. $p^2=m_1^2$ in
the integrand. In this way, even if propagators are fermions or
vectors, we have counted everything.}

\item The analytic formulas for tree amplitudes needed as input should
  be free of unphysical singularities involving the loop momentum $p$,
  so that the form of (\ref{I-inte}) is apparent. This is especially
  important when using on-shell recursion relations to derive
 tree-level amplitudes. Using Feynman diagrams or Berends-Giele
 recursion \cite{Berends:1987me}  to get
  tree-level expressions automatically circumvents this problem.

\item With our formulas, we can calculate any particular
coefficient directly without reference to other coefficients.

\item Our formulas can easily be used to obtain the 4-dimensional part of the
coefficients only,\footnote{That is to say, neglecting possible rational terms
which can be calculated by other methods, for example by the
recursive techniques of \cite{Bern:2005hs,Bern:2005ji,Bern:2005cq,Berger:2006ci,Berger:2006vq}
or the  specialized diagrammatic
reductions of \cite{Xiao:2006vr,Binoth:2006hk}.}  by taking the
limit $u \to 0$ in (\ref{RQ}). However, to be sure that all
intermediate formulas will be well-defined, it is safest to take
this limit at the end of the calculation. If we do wish to set $u\to 0$
at the beginning beginning, some care must be taken, as discussed in
Section \ref{6tri}.

\item Our formulas work for any $n$, although for renormalizable
theories we will have $n\leq 2$. But if we consider (super)gravity
or use a bad gauge choice, then we would have $n> 2$.

\item In spinor notation, we will find factors of the form
  $\gb{a|p|b}$ in the numerator.
This can be rewritten as $-2p\cdot P$ with $P=\la_a\W \la_b$.  So
$P_i$ can take complex values in (\ref{RQ}).

\end{itemize}

The formulas are presented in Section \ref{formulas}.  In Section
\ref{6g}, we present an example of a 4-dimensional  unitarity cut in
one helicity configuration of the six-gluon amplitude.  In Section
\ref{4g}, we give examples from the four-gluon amplitude but keep
the full $d$-dimensional dependence.  In Section \ref{discussion},
we discuss future applications and comparisons to other techniques.
Appendix \ref{spinorintegration} reviews the first steps in spinor
integration which form the basis of the derivation of our
coefficients.  Most details of the derivation of our formulas are presented in
Appendix \ref{derivation}. Appendix \ref{closed-triangle} contains
the formulas for triangle coefficients after the truncated series
expansion has been carried out explicitly, although direct the use of
the result in Section \ref{formulas} is likely to be simpler.

\section{\label{formulas}Coefficients of box, triangle and bubble integrals}

Here we give the results for the box, triangle, and bubble coefficients
in the unitarity cut defined by the momentum $K$, starting from the
integrand (\ref{I-inte}) without the prefactor $c$, and using the definitions (\ref{RQ}).
Our convention is that the  $n$-point scalar function is defined
by \footnote{We omit the prefactor $(-1)^{n+1}$ that is common
  elsewhere in the literature \cite{Bern:1992em,Bern:1993kr}.  
}
\bea I_n & = & i(4\pi)^{(4-2\eps)/2}\int {d^{4-2\eps} p \over
(2\pi)^{4-2\eps}}{1\over p^2 (p-K_1)^2 (p-K_1-K_2)^2...
(p-\sum_{j=1}^{n-1} K_j)^2}.~~~\label{n-scalar} \eea

The spinor notation we use here, which may differ from other
conventions,  is defined as follows.  For a four-vector $k_i$ satisfying $k_i^2=0$,
\bea \la_i\equiv u_+(k_i),~~~~\W \la_i\equiv u_-(k_i),
\eea
thus we have the following inner products:
\bea \vev{i~j}=\vev{i^-|j^+}=\O u_-(k_i)
u_+(k_j),~~~~[i~j]=[i^+|j^-]=\O u_+(k_i) u_-(k_j)\eea
Note that in this paper we use ``twistor'' sign conventions, so that
\bea 2k_i\cdot k_j= \vev{i~j}[i~j] \eea
which differs  from the standard QCD convention by a minus sign for
each spinor product $[i~j]$.
Our definitions implying the following relations:
\bea \gb{i|P|j} = \O u_{-}(k_i) \not P ~u_-(k_j),~~~~\vev{i|P_1
P_2|j}=\O u_{-}(k_i) \not P_1\not P_2 ~u_+(k_j)\eea

 The full $d$-dimensional amplitude will generically include pentagons in the basis.
 The identification of pentagon coefficients has already been described in \cite{Anastasiou:2006gt}.
  The operation occurs in the final integral over $u$.  Since our purpose here
  is to give the results of the 4-dimensional integration, we will not now
  comment any further on pentagons.  In cases involving massive species,
  tadpole integrals can also arise.  Unitarity methods cannot detect these.
  However, we expect that it will be possible to fix tadpole coefficients
  from other considerations, such as a heavy mass limit \cite{babis}.

\subsection{Box coefficients}

A box integral is identified by the two cut propagators plus two additional ones.
 Following the setup of the previous section, denote the two additional momenta
 associated to a box by $K_r$ and $K_s$.  Then, define the vectors $Q_r$ and
 $Q_s$ as in (\ref{RQ}).  From these two vectors, we construct two null
 vectors $P_{sr,1}$ and $P_{sr,2}$ as follows:\footnote{Here we see that
 the formula we give is ill-defined in special cases where $Q_r^2=0$.
 This case can arise if we have set $u$ to zero to find a 4-dimensional
 coefficient, and the external momentum $K_r$ is null.  However, there is
 no difficulty with the underlying method.  For a given box, $Q_r$ and $Q_s$
 can be exchanged.  Clearly, if both $K_r$ and $K_s$ are null, we
 can simply take $P_{sr,1}=Q_s, P_{sr,2}=Q_r$.
 In practice, this problem can always be avoided by keeping $u$ finite
 until the end of the calculation.}
\bea
\Delta_{sr} &=& (2Q_s \cdot Q_r)^2-4 Q_s^2 Q_r^2  \nonumber \\
P_{sr,1} &=& Q_s + \left( {-2Q_s \cdot Q_r + \sqrt{\Delta_{sr}}\over 2Q_r^2} \right) Q_r \nonumber \\
P_{sr,2} &=& Q_s + \left( {-2Q_s \cdot Q_r - \sqrt{\Delta_{sr}}\over 2Q_r^2} \right) Q_r~~~~\Label{box-null}
\eea
Then, the box coefficient with momenta $K,K_r,K_s$ is given by
\bea C[Q_r,Q_s,K] & = & {(K^2)^{2+n}\over 2}\left({\prod_{j=1}^{k+n}
\gb{P_{sr,1}|R_j |P_{sr,2}}\over \gb{P_{sr,1}|K
|P_{sr,2}}^{n+2}\prod_{t=1,t\neq i,j}^k \gb{P_{sr,1}|Q_t
|P_{sr,2}}}+ \{P_{sr,1}\leftrightarrow P_{sr,2}\}
\right).~~\Label{box-exp}\eea
%

\subsection{Triangle coefficients}

A triangle integral is identified by the two cut propagators plus one additional one.
 If the additional momentum variable is  $K_s$,  then define the vector $Q_s$ as
 in (\ref{RQ}).  Now construct two null vectors $P_{s,1}$ and $P_{s,2}$ as follows:

\bea
\Delta_{s} &=& (2Q_s \cdot K)^2-4 Q_s^2 K^2 \nonumber \\
P_{s,1} &=& Q_s + \left({-2Q_s \cdot K + \sqrt{\Delta_{s}}\over 2K^2} \right) K
\nonumber \\
P_{s,2} &=& Q_s + \left({-2Q_s \cdot K - \sqrt{\Delta_{s}}\over 2K^2} \right) K
~~~~\Label{tri-null}
\eea
Then, the triangle coefficient with momenta $K,K_s$ is given by
\bea C[Q_s,K] & = & { (K^2)^{1+n}\over
2}\frac{1}{(\sqrt{\Delta_s})^{n+1}}\frac{1}{(n+1)!
\vev{P_{s,1}~P_{s,2}}^{n+1}} \nonumber
\\ & & \times \frac{d^{n+1}}{d\tau^{n+1}}\left.\left({\prod_{j=1}^{k+n}
\vev{P_{s,1}-\tau P_{s,2} |R_j Q_s|P_{s,1}-\tau P_{s,2}}\over
\prod_{t=1,t\neq s}^k \vev{P_{s,1}-\tau P_{s,2}|Q_t Q_s
|P_{s,1}-\tau P_{s,2}}} + \{P_{s,1}\leftrightarrow
P_{s,2}\}\right)\right|_{\tau=0}.~~~~~\Label{tri-exp}\eea
In practice, the multiple derivative is easy to perform in a
symbolic manipulation program, either analytically or numerically,
and we believe this is an
efficient presentation of the coefficient. However, we have found
closed expressions, and these are given in Appendix
\ref{closed-triangle} for $n\leq 2$.

If $n \leq -2$, then the coefficient is simply zero.

\subsection{Bubble coefficients}

Every unitarity cut singles out a unique bubble integral.  However,
in our derivation, bubble and triangle integrals are related, so the
following formulas still require the quantities defined in
(\ref{tri-null}).  We further introduce two arbitrary {\sl real}
null vectors,\footnote{The reality condition is important.  These vectors
  should be physical momenta of massless particles.} $\eta$ and $\tilde\eta$,
and their associated spinors. These null vectors must, however, be
chosen generically: they should not coincide with other momentum
variables.

The coefficient of the bubble integral with momentum $K$ is given by
\bea
 C[K] = (K^2)^{1+n} \sum_{q=0}^n {(-1)^q\over q!} {d^q \over
ds^q}\left.\left( {\cal B}_{n,n-q}^{(0)}(s)+\sum_{r=1}^k\sum_{a=q}^n
\left({\cal B}_{n,n-a}^{(r;a-q;1)}(s)-{\cal
B}_{n,n-a}^{(r;a-q;2)}(s)\right)\right)\right|_{s=0},~~~~~\Label{bub-exp}
\eea
where
\bea {\cal B}_{n,t}^{(0)}(s)\equiv {d^n\over d\tau^n}\left.\left(
{1 \over n! [\eta|\W \eta K|\eta]^{n}}  {(2\eta\cdot
K)^{t+1} \over (t+1) (K^2)^{t+1}}{\prod_{j=1}^{n+k} \vev{\ell|R_j
(K+s\eta)|\ell}\over \vev{\ell~\eta}^{n+1} \prod_{p=1}^k \vev{\ell|
Q_p(K+s\eta)|\ell}}|_{\ket{\ell}\to |K-\tau \W \eta|\eta]
}\right)\right|_{\tau= 0},~~~\Label{cal-B-0}\eea
\bea & & {\cal B}_{n,t}^{(r;b;1)}(s)  \equiv  {(-1)^{b+1}\over
 b! \sqrt{\Delta_r}^{b+1} \vev{P_{r,1}~P_{r,2}}^b}{d^b \over d\tau^{b}}
\left({1\over (t+1)} {\gb{P_{r,1}-\tau P_{r,2}|\eta|P_{r,1}}^{t+1}\over
\gb{P_{r,1}-\tau P_{r,2}|K|P_{r,1}}^{t+1}}\right. \nonumber \\ & & \times \left.\left.
{\vev{P_{r,1}-\tau P_{r,2}|Q_r \eta|P_{r,1}-\tau P_{r,2}}^{b} \prod_{j=1}^{n+k} \vev{P_{r,1}-\tau P_{r,2}|R_j
(K+s\eta)|P_{r,1}-\tau P_{r,2}}\over \vev{P_{r,1}-\tau P_{r,2}|\eta K|P_{r,1}-\tau P_{r,2}}^{n+1} \prod_{p=1,p\neq
r}^k \vev{P_{r,1}-\tau P_{r,2}|
Q_p(K+s\eta)|P_{r,1}-\tau P_{r,2}}}\right)\right|_{\tau=0},~~~\Label{cal-B-r-1}\eea
\bea & & {\cal B}_{n,t}^{(r;b;2)}(s)  \equiv  {(-1)^{b+1}\over
 b! \sqrt{\Delta_r}^{b+1} \vev{P_{r,1}~P_{r,2}}^{b}}{d^{b} \over d\tau^{b}}
\left({1\over (t+1)} {\gb{P_{r,2}-\tau P_{r,1}|\eta|P_{r,2}}^{t+1}\over
\gb{P_{r,2}-\tau P_{r,1}|K|P_{r,2}}^{t+1}}\right. \nonumber \\ & & \times \left.\left.
{\vev{P_{r,2}-\tau P_{r,1}|Q_r \eta|P_{r,2}-\tau P_{r,1}}^{b} \prod_{j=1}^{n+k} \vev{P_{r,2}-\tau P_{r,1}|R_j
(K+s\eta)|P_{r,2}-\tau P_{r,1}}\over \vev{P_{r,2}-\tau P_{r,1}|\eta K|P_{r,2}-\tau P_{r,1}}^{n+1} \prod_{p=1,p\neq
r}^k \vev{P_{r,2}-\tau P_{r,1}|
Q_p(K+s\eta)|P_{r,2}-\tau P_{r,1}}}\right)\right|_{\tau=0}.~~~\Label{cal-B-r-2}\eea

Before ending this section, we want to make an additional remark. The
derivation of these formulas, given in Appendix \ref{derivation}, involved reducing the degree of $\W\la$  in both
numerators and denominators. However, we
could just as well choose to reduce the degree of
 $\la$ instead. In this case we would get formulas with the following
replacement: $\ket{\star}\to |\star]$ and $|\star]\to \ket{\star}$.
These two sets of  formulas are equivalent to each other in the case
$u\neq 0$. But if we naively set $u=0$ from the beginning, it is possible
that one of the two sets of formulas will break down.
Such an example will be seen in Section \ref{6tri}.

\section{\label{6g}An example from the six-gluon amplitude}

In this section we  test our formulas by computing some
coefficients from a one-loop partial amplitude with six external
gluons and an adjoint scalar circulating in the loop.  These
contribute to the full six-gluon amplitude in the spinor-helicity
formalism in
  the context of the supersymmetric decomposition
\cite{Bern:1993mq,Bern:1994zx,Bern:1994cg}. The box coefficient was
first computed in \cite{Bidder:2005ri}, and the bubble coefficient
was first computed in \cite{Britto:2006sj}.  The
two-mass-triangle coefficients have not appeared in this form
before, because it is possible to modify the basis and eliminate the
corresponding integrals, as described in
\cite{Britto:2005ha}.\footnote{In practice
  it is probably best to use the modified basis, because it is less divergent.}  In
\cite{Britto:2005ha} it was shown that for gluon amplitudes, these coefficients are
constrained by IR and UV divergences, and we use that relation here as a
consistency check.

Here we set the dimensional parameter $u$ to zero from the start in
order to work with simpler expressions.  As
described above, this simplification requires some care, and in fact we will see the
consequences when we derive the triangle coefficients.

We choose the unitarity cut of the momentum $K \equiv k_4+k_5+k_6$ in the
helicity configuration
$(1^-2^-3^+4^-5^+6^+)$.
The cut integral is
\bea
C_{123} &=& \int d\mu~
2 A(\ell_1^-,1^-,2^-,3^+,\ell_2^+)A((-\ell_2)^-,4^-,5^+,6^+,(-\ell_1)^+)
\nonumber \\ &= &
 {2 \over s_{456}[1~2][2~3]\vev{4~5}\vev{5~6}} \int d\mu~
{\vev{4~\ell_1}^2\vev{4~\ell_2}[3~\ell_1]^2[3~\ell_2]  \over
\vev{6~\ell_1}[\ell_1~1]}
\nonumber \\ &=&  -{2 \over s_{456} \cb{1~2}\cb{2~3}\vev{4~5}\vev{5~6}}
\int d\mu~{\gb{1|\ell|6} \gb{4|\ell|3}^3 \over (\ell-k_6)^2 (\ell+k_1)^2}
~~~\Label{sixcut}
\\ & & + {2 \gb{4|K|3}  \over s_{456} \cb{1~2}\cb{2~3}\vev{4~5}\vev{5~6}}
\int d\mu~{\gb{1|\ell|6} \gb{4|\ell|3}^2 \over (\ell-k_6)^2 (\ell+k_1)^2}
\nonumber
\eea
So we have two terms, each with $k=2$, and
\bean
K_1=k_6, ~~~~K_2=-k_1,
\eean
so
\bean
Q_1=-k_6, ~~~~Q_2=k_1.
\eean
  In the first term $m=4$, and in the second term $m=3$.  We have
\bean R_1 &=& -\lambda_1\tilde\lambda_6, ~~~~ R_2 = R_3 = R_4 =
-\lambda_4\tilde\lambda_3. \eean
%
%

\subsection{Box coefficient}

Since $k=2$, we see immediately that there can be only one
nonvanishing box coefficient in this cut. We compute the null
vectors $P_{sr,1}$ and $P_{sr,2}$ from the definitions in
(\ref{box-null}), and define the associated spinors as
follows.\footnote{We could just as well use \bean P_{21,1}=Q_2=k_1,
& \ket{P_{21,1}}=\ket{1}, & |P_{21,1}]=|1],\\
  P_{21,2}=Q_1=-k_6,
& \ket{P_{21,2}}=\ket{6}, & |P_{21,2}]=-|6].\eean }
\bean
\begin{array}{l l l}
  P_{12,1}=Q_1=-k_6
 ~~~~~& \ket{P_{12,1}}=\ket{6}
~~~~~~& |P_{12,1}]=-|6] \\
  P_{12,2}=Q_2=k_1
& \ket{P_{12,2}}=\ket{1}
& |P_{12,2}]=|1] \\
\end{array}
\eean

Applying (\ref{box-exp}) to the expressions under the integral signs
in (\ref{sixcut}),
we get
\bean C[Q_1,Q_2,K] & = &  {s_{456}^{2+n}\over 2}\left( {
(-1)^{n+2} s_{61} \vev{6~4}^{n+1} \cb{3~1}^{n+1} \over
\gb{6|K |1}^{n+2}}\right) \eean
Now attach the prefactors for each of the two terms.
For the first term with $n=2$:
\bean  -{2 \over s_{456}[1~2][2~3]\vev{4~5}\vev{5~6}} \times
C[Q_1,Q_2,K] & = &  -{ s_{456}^3 s_{61}\vev{6~4}^{3} \cb{3~1}^{3}\over   [1~2][2~3]\vev{4~5}\vev{5~6}\gb{6|K
|1}^{4}}
\eean
For the second term with $n=1$:
\bean  {2\gb{4|K|3} \over s_{456}[1~2][2~3]\vev{4~5}\vev{5~6}}
\times C[Q_1,Q_2,K] & = &  - {s_{456}^{2} s_{61} \vev{6~4}^{2}
  \cb{3~1}^{2}\gb{4|K|3} \over [1~2][2~3]\vev{4~5}\vev{5~6}\gb{6|K
|1}^{3}}
\eean
So the total coefficient of the box $(1|23|45|6)$, for the scalar contribution, is
\bea
-{s_{456}^2 s_{61} \vev{6~4}^{2} \cb{3~1}^{2}\gb{6|K|3}\gb{4|K|1}
\over [1~2][2~3]\vev{4~5}\vev{5~6}\gb{6|K|1}^{4}}~~~\Label{6box}
\eea
This  agrees with the expression in \cite{Bidder:2005ri}  when we
incorporate the usual factor of $2/(s_{456} s_{61})$.

\subsection{\label{6tri}Triangle coefficients}

Since $k=2$, we see immediately that there can be only two
nonvanishing triangle coefficients in this cut.

For $Q_1$:
\bean
\begin{array}{lll}
 \sqrt{\Delta_1} = -s_{456}+s_{45}  & & \\
  P_{1,1} = -k_6 &  \ket{P_{1,1}}=\ket{6} & |P_{1,1}]=-|6] \\
 P_{1,2}
={s_{456}-s_{45}\over s_{456}}(k_4+k_5)-{s_{45}\over s_{456}}k_6
~~~~&  \ket{P_{1,2}}={K|6]\over s_{456}}
~~~~&  |P_{1,2}] = K\ket{6}
\end{array}
\eean

For $Q_2$:
\bean
\begin{array}{lll}
 \sqrt{\Delta_2}= -s_{456}+s_{23} & &  \\
P_{2,1}  = k_1
& \ket{P_{2,1}}=\ket{1} & |P_{2,1}]=|1]\\
P_{2,2}
={s_{23}\over s_{456}} k_1-{s_{456}-s_{23}\over s_{456}}(k_2+k_3)
~~~~& \ket{P_{2,2}}= {K|1]\over s_{456}}
~~~~& |P_{2,2}]= -K\ket{1}
\end{array}
\eean
Let us first consider the triangle $(1|23|456)$, with momenta $K$
and $K_2$ ($Q_2$). With the identity
\bean
\frac{d^{n+1}}{d\tau^{n+1}}\left.\left(
{(a-\tau b)^{n+1}(-\tau)^{n+2}
\over c-\tau d}
+ {(b-\tau a)^{n+1}
\over
d-\tau c}
\right)\right|_{\tau\to 0}
= {(n+1)!(bc-ad)^{n+1} \over d^{n+2}}
\eean
we see that the
formula (\ref{tri-exp}) for triangle coefficients becomes
\bean C[Q_2,K] & = & { s_{456}^{1+n}\over 2}
{\gb{1|K|1}\vev{4~6}^{n+1}\cb{3~ 1}^{n+1} \over  \gb{6|K|1}^{n+2}}
\eean
Adding the $n=1$ and $n=2$ contributions and attaching the prefactors,
we find that the total coefficient is
\bea
& &
 -{2 \over s_{456}[1~2][2~3]\vev{4~5}\vev{5~6}}
\left(
 { s_{456}^{3}\over
2}
{\gb{1|K|1}\vev{4~6}^{3}\cb{3~ 1}^{3}
\over  \gb{6|K|1}^{4}}
\right)
\nonumber
\\ & & + {2\gb{4|K|3} \over s_{456}[1~2][2~3]\vev{4~5}\vev{5~6}}
\left(
 { s_{456}^{2}\over
2}
{\gb{1|K|1}\vev{4~6}^{2}\cb{3~ 1}^{2}
\over  \gb{6|K|1}^{3}}
\right)
\nonumber
\\ & &
= {s_{456} \gb{1|K|1}\gb{4|K|1}\gb{6|K|3}\cb{3~ 1}^{2}\vev{4~6}^{2}
\over [1~2][2~3]\vev{4~5}\vev{5~6}\gb{6|K|1}^{4}}
~~~\Label{first-6tri}
 \eea

Now consider the triangle $(123|45|6)$, with momenta $K$ and $K_1$
($Q_1$). A naive application of the formula (\ref{tri-exp}) in the
limit $u \to 0$ gives
\bean
 C[Q_1,K] & = & 0
\eean
because of the $R_1Q_1$ contraction in the numerator's product. But
the triangle coefficient does not actually vanish!  This is clear,
because this triangle is related by conjugation and label
permutation to the previous one.  This is the degenerate case that we
 discussed at the end of the previous section. The reason is clear.
From (\ref{sixcut}), we see that for $Q_2$, the pole is $[\ell~1]$,
while for $Q_1$ it is $\vev{\ell~6}$. Since the formula given in
previous section is obtained by writing total derivative in $[d\W
\a~\partial_{\W \la}]$, they are  not suitable for pole
$\vev{\ell~6}$. To deal with it we need to use the conjugate formula where
we replace
 $\ket{\star}\to |\star]$ and $|\star]\to
\ket{\star}$, i.e., writing a total derivative of the form $\vev{d
\a~\partial_{ \la}}$. This will be a general rule in all
4-dimensional calculations with null poles. We want to emphasize
that  this situation will not arise if we keep $u \neq 0$ until the
end.

After clarifying the subtle point we can continue our calculation.
Either by taking the conjugate of the formula (\ref{bub-exp}), which is
\bean  {  s_{123}^{n+1}\over
2(n+1)!(\sqrt{\Delta_1})^{n+1}[P_1~P_2]^{n+1}}{d^{n+1}\over
d\tau^{n+1}}\left( {[P_1-\tau P_2|R_1 Q_1|P_1-\tau P_2][P_1-\tau
P_2|R_2 Q_1|P_1-\tau P_2]^{n+1}\over [P_1-\tau P_2|Q_2 Q_1|P_1-\tau
P_2]}+\{P_1\leftrightarrow P_2\}\right),
\eean
or by directly applying the relabeling and conjugation to
(\ref{first-6tri}), we find

\bean  C[Q_1,K] & =& - {
s_{456}^{n+1}\gb{6|K|6}\vev{6~4}^{n+1}[1~3]^{n+1}\over 2
\gb{6|K|1}^{n+2}}.\eean

Adding the two terms from $n=2$ and $n=1$, with prefactors,  we get
\bea & & -{s_{456} \gb{6|K|6} \vev{6~4}^2[1~3]^2\over
[1~2][2~3]\vev{4~5}\vev{5~6}}(-s_{456}\vev{6~4}[1~3]+\gb{4|K|3}\gb{6|K|1})
\nonumber
\\
& =& -{s_{456} \gb{6|K|6}
\vev{6~4}^2[1~3]^2\gb{4|K|1}\gb{6|K|3}\over
[1~2][2~3]\vev{4~5}\vev{5~6}\gb{6|K|1}^4}~~~\Label{second-6tri}
\eea

~\\{\bf Consistency check:\\}

These particular triangle coefficients have not been isolated before,
because two-mass triangles disappear in the modified integral basis proposed in
\cite{Britto:2005ha}.
We can now perform a consistency check
based on the same identity
that allowed the basis to be modified.  Consider all the contributions to the
divergence $(-s)^{-\eps}$, where $s=K^2$.  In these example, there are exactly these
two 2-mass triangles plus the single box from the previous subsection.
The condition expressing the vanishing of this divergence is \footnote{The
  relative sign between box and triangle terms comes because our sign
  conventions for the master integrals (\ref{n-scalar}) differ from those of \cite{Britto:2005ha}.}
\bea
0 & = &   c^{2m~h}_4 {2 \over s t} - \sum c^{2m}_3(s,t) {1 \over (-s)-(-t)}
\\ &=  &  c^{2m~h}_4 {2 \over s_{456} s_{61}}
-  c_{[1|23|456]} {1 \over -s_{456}+s_{23}} -  c_{[123|45|6]} {1 \over
  -s_{456}+s_{45}} \nonumber
\\ &=  &  c^{2m~h}_4 {2 \over  s_{456}s_{61}}
+  c_{[1|23|456]} {1 \over \gb{1|K|1}}-  c_{[1|23|456]} {1 \over
  \gb{6|K|6}}. \nonumber
\eea
It is easy to see that this identity is satisfied by our coefficients
given in (\ref{6box}),(\ref{first-6tri}),(\ref{second-6tri}).

\subsection{Bubble coefficient}

Let us choose  $\eta=3$ and
$\tilde\eta=4$.   This choice gives somewhat simpler formulas; for
example, $ {\cal B}_{n,n-q}^{(0)}(s)$ and ${\cal B}_{n,n-a}^{(2;a-q;1)}(s)$ are identically zero, because there is a
sufficiently high power of $\tau$ inside the derivative.
We find
\bean
 {\cal B}_{n,n-q}^{(0)}(s)= 0
\eean
\bean {\cal B}_{n,n-a}^{(1;a-q;1)}(s) & = & {(-1)^{a}  \cb{6~3}^{a-q}\over
\gb{6|K|6}^{n-q+2} (a-q)! }{d^{a-q} \over d\tau^{a-q}}
\left({  \cb{3~6}^{n-a+1} \over s_{456}^{n+1-q} (n-a+1)}
\right. \nonumber \\ & & \times \left.\left.
{
\tau^{a-q}
(s_{456} \tgb{6|K+s k_3|6} - \tau s \cb{6|3|K|6})
(s_{456} \vev{4~6} - \tau \gb{4|K|6})^{n+1}
\over
(s_{456} \vev{3~6} - \tau \gb{3|K|6})^{q}
(s_{456} \tgb{1|K+s k_3|6} - \tau \cb{1|K+s k_3|K|6})
}\right)\right|_{\tau =0}
\eean
\bean
 {\cal B}_{n,n-a}^{(1;a-q;2)}(s) & = &  {(-1)^{q} \cb{6~3}^{a-q}\over
\gb{6|K|6}^{n-q+2} (a-q)! }{d^{a-q} \over d\tau^{a-q}}
\left({  \tgb{3|K|6}^{n-a+1} \over  s_{456}^{n+1-q} (n-a+1)}
\right. \nonumber \\ & & \times \left.\left.
{
(s \cb{6|3|K|6} - \tau s_{456} \tgb{6|K+s k_3|6})
(\gb{4|K|6} - \tau s_{456} \vev{4~6})^{n+1}
\over
(\gb{3|K|6} - \tau s_{456} \vev{3~6})^{q}
(\cb{1|K+s k_3|K|6} - \tau s_{456} \tgb{1|K+s k_3|6})
}\right)\right|_{\tau =0}
\eean
\bean {\cal B}_{n,n-a}^{(2;a-q;1)}(s)
&=& 0
\eean
\bean
 {\cal B}_{n,n-a}^{(2;a-q;2)}(s) & = &
{(-1)^{q}  \cb{1~3}^{a-q} \over
\gb{1|K|1}^{n-q+1} (a-q)! }{d^{a-q} \over d\tau^{a-q}}
\left({  \tgb{3|K|1}^{n-a+1} \over s_{456}^{n+1-q}  (n-a+1)}
\right. \nonumber \\ & & \times \left.\left.
{
(\gb{4|K|1} - \tau s_{456} \vev{4~1})^{n+1}
\over
(\gb{3|K|1} - \tau s_{456} \vev{3~1})^{q}
(\gb{6|K|1} - \tau s_{456} \vev{6~1})
}\right)\right|_{\tau =0}
\eean
We then substitute these expressions into (\ref{bub-exp}) and attach
the prefactors. For $ {\cal B}_{n,n-a}^{(2;a-q;2)}(s)$, in fact only
the $q=0$ contributions matter, because the $s$-dependence has
dropped out with our choice of $\ket{\eta}=\ket{3}$. We have checked
numerically that the result agrees with the corresponding result
derived by the technique of \cite{Britto:2006sj}.\footnote{We were
  unable to confirm numerically the printed value of the corresponding
coefficient in \cite{Britto:2006sj}, so we repeated the calculation.}

\section{\label{4g}A $d$-dimensional example: four gluons}

In this section we illustrate the use of the  formulas in Section
\ref{formulas} in the case of
four gluons with a scalar propagating in the loop.
These amplitudes were first given in
 \cite{Bern:1995db}.  Our notation and presentation here are more
 similar to
\cite{Brandhuber:2005jw} and especially \cite{Anastasiou:2006gt},
where these amplitudes were derived by newer techniques.
Here we have verified that our results reproduce those in the
literature.\footnote{The $-+-+$ amplitude was given the wrong overall
  sign in equation (4.31) of
  \cite{Anastasiou:2006gt}.}
We stop just before
 the final integral over
$u$, which could in general be done by the techniques of
\cite{Anastasiou:2006jv,Anastasiou:2006gt}, developed in the context of
$d$-dimensional unitarity \cite{vanNeerven:1985xr,Bern:1995db,Bern:1996je,Bern:1996ja,Brandhuber:2005jw}.  Note therefore that the
labels of ``box, triangle, bubble'' are used in the $d$-dimensional sense.
The four-gluon amplitude is a nice test of our formulas, because this
simple case is where they are most likely to break down, for example
by a bad choice of $\eta$, as we shall see in the last configuration.  We consider
three of the four independent helicity configurations, since the
fourth adds no new features.

For ease of presentation, we make use of the variable $z$ as given in (\ref{z-u}).

\subsection{$(1^+,2^+,3^+,4^+)$ }

The simplest helicity configuration is $(++++)$. The integrand for the
cut
$K=K_{12}$ is
\bean {2\mu^4[1~2][3~4] \over \braket{1~2}\braket{3~4}}{1 \over
(p-k_1)^2 (p+k_4)^2}\eean
From this, by comparing with general formula (\ref{I-inte}) we have
$m=0$, $k=2$, $K_1=k_1, K_2=-k_4$, thus $n\equiv m-k=-2$, so there
are neither triangle nor bubble contributions. There is only one box
coefficient. There are no vectors $P_i$.  The expression inside the
parentheses in  (\ref{box-exp}) thus degenerates to 1, so we find
\bea {2\mu^4[1~2][3~4] \over \braket{1~2}\braket{3~4}} {1\over
2}(1+1)= {2\mu^4[1~2][3~4] \over \braket{1~2}\braket{3~4}}. \eea
%

\subsection{$(1^-,2^+, 3^+,4^+)$ }
For this case we have two cuts $C_{12}$ and $C_{41}$. These
two cuts are related to each other by symmetry, so we focus on cut
$C_{41}$. The integrand is
 \bean  -{u[2~3]\over 2\vev{2~3}}{ \gb{1|p|4}^2\over (p-k_4)(p+k_3)^2} \eean
For this case we have $m=k=2$ so $n=m-k=0$, $K_1=k_4, K_2=-k_3$,
thus
\bean & & P=\la_1 \W\la_4,~~~~~R=R_1=R_2=
-(1-2z)\la_1 \W\la_4  \nonumber \\
& &  Q_1= -z k_1-(1-z)
k_4,~~~~~~Q_2= (1-z) k_3+z k_2
\eean

{\bf Box:}
Using
\bean Q_1^2 & = & Q_2^2 = {u\over 4} K_{41}^2,~~~~
2Q_1\cdot Q_2  =  -K_{12}^2 +{u\over 2}(K_{12}^2-K_{13}^2) \\
\Delta_{12} & = & s^2(1-u)(1+u {t\over
s}),~~~s=K_{12}^2,~t=K_{13}^2,\eean
 we have
\bea -{u[2~3] \over 2 \vev{2~3} }{ \tgb{4|3|1}^2 (2+A u)\over 2
s_{13}^2} {s_{41}^2\over 2}= -{s_{41} [2~3]^2[4~3]^2 u(2+A u)\over 8
[1~3]^2},~~~~~~A={s_{13} \over s_{12}}.
\eea

{\bf Triangle:}

Formally, there are two triangles identified by
the momenta $K_1$ and $K_2$.
 In this special example with only four
external particles, they are, in fact, the {\it same} triangle. So we
will need to add these two contributions to the final coefficient.

Let us start with $K_1$. From (\ref{tri-null}), we find
\bean
\Delta_1  =  (1-2z)^2 s_{14}^2, ~~~~~P_{1,1}=(1-2z) k_1, ~~~~~P_{1,2}=-(1-2z)
k_4.
\eean
 For the spinors, we choose
\bean \ket{P_{1,1}}=\ket{1},~~|P_{1,1}]=(1-2z)|1],~~
\ket{P_{1,2}}=\ket{4},~~|P_{1,2}]=-(1-2z)|4].
\eean
Since $n=0$, using (\ref{tri-exp}) we get
\bean  -{  (1-2z)\over 2 \vev{4~1}}{d\over d\tau} \left(
{z^2\vev{4~1}^2[4~1]^2\vev{1~4}^2\over \vev{k_4-\tau k_1|Q_2
Q_1|k_4-\tau k_1}}+{z^2 \tau^4 \vev{4~1}^2[4~1]^2\vev{1~4}^2\over
\vev{k_1-\tau k_4|Q_2 Q_1|k_1-\tau k_4}}\right)
 & = &-{s_{12}s_{41}^2\over 2 \gb{4|3|1}^2}\eean
With the prefactor included, we have
\bean -{u[2~3] \over 2 \vev{2~3}}C[Q_1,K_{41}]={u s_{41} [2~3]^2[3~4]^2\over 4 s_{12}
[3~1]^2}
\eean

For the triangle with $K_2$ we have
\bean \Delta_2 & = &(1-2z)^2 s_{23}^2,
~~~P_{2,1}=-(1-2z) k_2,~~~P_{2,2}=(1-2z) k_3.
\eean
Similar calculations give
\bean -{u[2~3] \over 2 \vev{2~3}}C[Q_2,K_{41}]=
-{u[2~3]^2[4~3]^2\over 4 s_{23} [3~1]^2}\left(
{s_{13}^2\over s_{12}}-2 s_{13}-s_{12}\right)
.\eean
Adding these two contributions together,  we find that the triangle
coefficient is
\bea -{u[2~3]^2[4~3]^2  s_{12}(s_{41}-s_{13})\over 2 s_{23} s_{12}
[3~1]^2}.
\eea

{\bf Bubble:}
The formula (\ref{bub-exp}) reduces to
\bean C[K] & = & {[\eta|R K|\eta]^2\over [\eta|Q_1
K|\eta][\eta|Q_2 K|\eta]} -\sum_{r=1}^2\left\{{K^2\over
\sqrt{\Delta_r}} {\gb{P_{r,1}|\eta|P_{r,1}}\over
\gb{P_{r,1}|K|P_{r,1}}}{\gb{P_{r,1}|R|P_{r,2}}^2\over
\gb{P_{r,1}|\eta|P_{r,2}}\gb{P_{r,1}|Q_2|P_{r,2}}}-\{P_{r,1}\leftrightarrow
P_{r,2}\}\right\}.\eean
Taking $\eta=k_4$ it is easy to see that both the first term and
the $r=1$ terms are zero, so we are left with the $r=2$ terms only.  The result
is
\bean -{[4~3]^2 \over [3~1]^2}{s_{13}\over s_{41}}\left(
1-{s_{13}\over s_{12}}\right).
\eean
With the prefactor included, we find that the bubble coefficient is
\bea   -{u[2~3] \over 2 \vev{2~3}}C[K] & = &{ u  s_{13}(s_{12}-s_{13})[2~3]^2[3~4]^2\over 2 s_{41}^2
s_{12} [1~3]^2}.
\eea
%

\subsection{$(1^-,2^+,3^-,4^+)$}

The integrand for the cut $K_{41}$ is given by
\bean {2 \gb{1|\ell|4}^2 \gb{3|\ell|2}^2\over s_{41}^2
((\ell-k_4)^2-\mu^2) ((\ell+k_3)^2-\mu^2)} \eean
For this case we have  $m=4$, $k=2$, so $n=m-k=2$, $K_1=k_4,
K_2=-k_3$, thus
\bean & & R_1= R_2=-(1-2z)\la_1 \W\la_4,~~~~~R_3=R_4=-(1-2z) \la_3 \W\la_2. \nonumber \\
& &  Q_1= -z k_1-(1-z) k_4, ~~~~~~~~~ Q_2=
(1-z) k_3+z k_2.
\eean

{\bf Box:} By now it is straightforward to find that the coefficient is
\bea {2 \over s_{41}^2} C[Q_1, Q_2,K] ={ \vev{1~3}^2 s_{41}^2 ( 8
s_{12}^2+ 8 s_{12} s_{13} u+ s_{13}^2 u^2)\over 8 \vev{2~4}^2
s_{13}^2}.
\eea

{\bf Triangle:}
For the first triangle, with $K_1$, we have
\bean C[Q_1,K] & = &  {1 \over 12 (1-2z)^3
 \vev{1~4}^3}{d^3\over d\tau^3}\left( {\vev{k_1-\tau k_4|R_1
Q_1|k_1-\tau k_4}^2\vev{k_1-\tau k_4|R_3 Q_1|k_1-\tau k_4}^2\over
\vev{k_1-\tau k_4|Q_2 Q_1|k_1-\tau k_4}} + \{ k_1\leftrightarrow
k_4\}\right). \eean
In this case, the factor $\vev{k_1-\tau k_4|R_1
Q_1|k_1-\tau k_4}$ is proportional to $\tau^2$, so the contribution of first term is zero and we
have
\bean {2\over s_{41}^2}C[Q_1,K]& =&  -{ \vev{1~3}^2 s_{13}\over 2 \vev{2~4}^2} { (1+\W
A)^2(2+ \W A u)\over \W A^3},~~~\W A={s_{13} \over s_{12}}
\eean
By symmetry, for the triangle with momentum $K_2$ we just need
to do exchange the labels $1\leftrightarrow 2, 3\leftrightarrow 4$, so the
final coefficient is just twice what is written above, namely
\bea  -{ \vev{1~3}^2 s_{13}\over  \vev{2~4}^2} { (1+\W A)^2(2+ \W
A u)\over \W A^3},~~~\W A={s_{13} \over s_{12}}.
\eea

{\bf Bubble:}
To present this example analytically, we choose $\eta=k_4$ rather than
a more generic value.
However, we do run into a problem then, because of an accidental
degeneracy of poles.  This problem could have been avoided by choosing a
generic $\eta$, but for analytic purposes, we considered this case
separately, and it is presented in Appendix \ref{specialcase}.  The
formula  (\ref{spe-Re-gen-n-1}) for the bubble coefficient now takes
modified input, as given in  (\ref{spe-cal-B-0}),
(\ref{spe-cal-B-r-1}) and (\ref{spe-cal-B-r-2}).

First, consider the term (\ref{spe-cal-B-0}).  Since $n=2$, there is a third
derivative in $\tau$.  Since the factor $[4|(K-\tau \W\eta) R_1(K+s
k_4)(K-\tau \W\eta)|4]$ is proportional to $\tau^2$, this contribution vanishes.
 Similarly, (\ref{spe-cal-B-r-1}) will be zero.  The only remaining part is the $r=2$
case, (\ref{spe-cal-B-r-2}), with
\bean
P_1=-(1-2z)k_2,~~~~ P_2= (1-2z) K_3, ~~~~\Delta_{r=2}=(1-2z)
s_{41}^2.
\eean
Let us discuss ${\cal B}_{2,t}^{(2;a;2)}(s)$ first.
Notice that
\bean  \vev{3-\tau 2|R_3 (K+s \eta)|3-\tau 2}  & = &(1-2z) \tau
\vev{2~3} s_{12}\O B ( - s +{\tau\over \O B}(1 +(1+s) \W A)). \eean
We see that to get a nonzero value of ${\cal B}_{2,t}^{(2;a;2)}(s)$ we
must have $a=2$, and more specifically,
\bean {\cal B}_{2,t}^{(2;a;2)}(s) & = & 0,~~~a=0,1 \\
{\cal B}_{2,t}^{(2;a=2;1)}(s) & = & -{(1-2z) [4~1]^2 \O B^{4}
\vev{1~2}^4 \over  s_{41}^{3}\vev{2~3}^2}{(-s_{12})^{t+1} \over
(t+1) s_{41}^{t+1}} {z^2 s^2 \over (1-2z)-zs)}\eean
Because the factor $s^2$ appears in ${\cal B}_{2,t}^{(2;a=2;1)}(s)$,
there is a nonzero contribution only when we take $q=2$ in the
derivative with respect to $s$.
But then $a-q=0$, and since ${\cal B}_{2,t}^{(2;a-q;2)}(s)$ is only
nonzero for $a-q=2$,
the contribution from this part is zero.

For ${\cal B}_{2,t}^{(2;a;1)}$ we have
\bean {\cal B}_{2,t}^{(2;a;1)}(s) & = & {\vev{1~3}^2\over
\vev{2~4}^2 \W A^2 s_{41}}{(-1)^a (1-2z)^{3-a} \W A^{t+1} \over a!
(t+1) (1+\W A)^{a+t+1}}  {d^a \over d\W \tau^a} \left( { (1-\W
\tau)^{t+1}(1-z+z \W A \W \tau )^a (s \W A \W \tau-1-(1+s) \W
A)^2\over (1-\W \tau)^{4-a} ((1-2z)-zs)}\right), \eean
where $\W \tau=\tau/B$, $B={\vev{2~4}/ \vev{3~4}}$, and $\W
A=s_{13}/s_{12}$.
Summing up $(t,a)=(2,0),(1,1),(0,2)$ with $s=0$, $(t,a)=(1,0),(0,1)$
with the first derivative of $s$, and $(t,a)=(0,0)$ with the second
derivative of  $s$, we finally find that the coefficient is
\bea {2 \over s_{41}^2 }C[K] & = & -{2\vev{1~3}^2 s_{13}\over \vev{2~4}^2  s_{41}} {(12 +3\W A(6+u)+
\W A^2(4+5 u))\over 12 \W A^2},~~~~~~\W A = {s_{13} \over s_{12}}.
\eea
%

\section{\label{discussion}Discussion}

Since the formalism described here is based on unitarity cuts of the
amplitude, it shares with other unitarity-based
approaches\footnote{For a review, see Section 4 of \cite{Bern:2007dw}.} the
property that the input required is simply a collection of
tree-level amplitudes.  These are manifestly gauge invariant and can
take quite compact forms.  By dealing with different cuts
separately, we can attack the problem in stages.

Furthermore, our formulas separate and identify the coefficients of
individual master integrals.  A single unitarity cut yields, directly
and separately,
the coefficients of all the master integrals with the same cut propagators.
Any single coefficient can be targeted individually, without the need
to first compute any others or additional spurious terms.

\subsection{Comparison with other approaches}

The reduction algorithm of Ossola, Papadopoulos and Pittau (OPP)
\cite{delAguila:2004nf,Ossola:2006us,Ossola:2007bb} produces coefficients through
algebraic operations at the integrand level, through recursive
solution of a set of algebraic equations.
In fact, our formulas given here are  the results of solving algebraic
equations in a
different style. In the  OPP method, several points of phase space are used, while
in our method, we differentiate at a single point of phase space.  The
derivative operator can be interpreted as an algebraic procedure as
applied to rational functions at a single point.

Just like our result, coefficients from OPP method can be fed into
the $d$-dimensional unitarity program as described in
\cite{Anastasiou:2006jv, Anastasiou:2006gt}.
 Alternatively, the algorithm may be
interpreted  numerically, and in fact such an implementation
has now been given \cite{Ossola:2007ax} (see also the procedure of
\cite{Ellis:2007br}).  We believe that the formulas of the present
paper are also well suited for numerical programming, and this will
be the subject of forthcoming work.

One final note on comparison to the OPP method is that our formulas
are valid for arbitrary values of $n$, in particular for $n>2$. Such
an extension has been mentioned within the OPP method, although details have
not been worked out.

An approach that is closer in spirit to ours was given by Forde in
\cite{Forde:2007mi}.  There, coefficients for boxes, triangles and
bubbles are given within the spinor formalism.  The foundation there
consists of generalized unitarity cuts, namely quadruple cuts for
boxes, triple cuts for triangles, and ordinary double cuts for
bubbles.  The motivation was to capitalize on the efficiency of
quadruple cuts for box coefficients, and also to be able to target
specific coefficients.  Forde's final formulas resemble ours in that
they are based on data from tree amplitudes and given in terms of a
coefficient in a series expansion of one variable for triangles, and
two variables for bubbles.  The formula for a bubble coefficient,
however, depends on tree amplitude input for all possible triple cuts, while ours
comes directly from the ordinary double cut (though still depending on
all possible momenta from a hypothetical third cut). If
the aim is to assemble an amplitude in its entirety, then of course
the complete tree-level amplitude input will be available anyway.

In \cite{Britto:2006fc}, we discussed the application of quadruple
cuts to box and pentagon integrals in $d$ dimensions. A
$d$-dimensional analysis of triple cuts for triangle integrals has
been given in \cite{Mastrolia:2006ki}.  Let us now briefly examine the
triple cut in the context of the present paper in order to make
contact with the result of \cite{Forde:2007mi}.

 With three delta functions in $d$ dimensions, we first use two of them
to set up the four-dimensional spinor integrand, as explained in
Appendix \ref{spinorintegration}.  After integrating over the variable
$t$, defined in (\ref{spinormeasure}), we arrive at the integral
\bean
 \int \vev{\ell~d\ell}[\ell~d\ell]~ G(\ell)~ \delta \left(
{K^2\gb{\ell|Q|\ell}\over \gb{\ell|K|\ell}}\right)
\eean
 Now we can use momenta $Q,K$ to construct
two null momenta as in (\ref{tri-null}) and expand our spinor
variables in the basis of their spinor components as follows:
\bean
 \ket{\ell}=\ket{P_1}+z\ket{P_2},~~~~~|\ell]=|P_1]+\O z|P_2]
\eean
Here $z$ is a complex number and $\O z$ is its conjugate.
With this substitution, we get
\bea \int dz~d\O z~ (2P_1\cdot P_2)~ G(z,\O z)~ \delta\left( K^2{
\gb{P_1|Q|P_1}+z\O z\gb{P_2|Q|P_2}\over \gb{P_1|Q|P_1}+z\O
z\gb{P_2|Q|P_2}}\right).
\eea
Now we can change to polar coordinates so that $z=r e^{i\theta}$ and $dzd\O z=  rdr d\theta$.
Furthermore, if we now define the new variable $t=e^{i\theta}$, we have
\bean
 dz d\O z= rdr \times {-i dt\over t} .
\eean
 The delta function depends only on $r$, so we can use it
to integrate over $r$. Then we are left with $t$ integration only.
From here, for example, it is easy to see the vanishing condition
given in  eq. (4.20) of \cite{Forde:2007mi}.

Furthermore, for box integrals, we have an extra propagator, so  the
general form of the integrand is ${1/( a+t b+t^{-1} c)}$. Only polynomial terms
correspond to the triangle contribution.

The parametrization we have used here is not exactly the one used by
Forde, but the central idea is the same and $t$ is the angle
variable for both triple cuts and double cuts.

\subsection{Prospects}

The most obvious and immediate application of our results will be
to the computation of complete one-loop amplitudes, as in
the example of Section \ref{4g}, where the $u$-dependent expressions for
coefficients are fed into the reduction formulas of
\cite{Anastasiou:2006jv, Anastasiou:2006gt} to give the final
$\eps$-dependent coefficients.  The four-momenta of the external
particles may be numerical at every step.  The reduction formulas
currently require analytic expressions in $u$.

Our formulas may also be specialized to the cut-constructible part of
the amplitude, as in the example of Section \ref{6g}, simply by
setting $u \to 0$ at the end of the calculation and interpreting the
formulas as exact coefficients of 4-dimensional master integrals.

Since several methods proposed in the literature are specialized for computing
either cut-constructible or rational components of an amplitude,
it would also be very interesting to specialize our formulas to
isolate the rational part of one-loop amplitudes.  This could be
done by studying the $\eps$-dependence of the reduction formulas
together with the $u$-dependence in the coefficients, in order to
focus precisely on the $\eps^0$ term in the final $\eps$-expansion
of the amplitude. We will return to this point in a future publication.

Finally, as we remarked in the introduction, our formulas apply to
amplitudes with massive or massless propagators.  In order to arrive
at complete amplitudes in the massive case, the master integrals
should be evaluated explicitly,\footnote{For a uniform mass, this has
  been done in \cite{Bern:1995db}.} and our
results will need to be supplemented with the contributions of tadpole
and massless bubble integrals.

\acknowledgments

We are grateful to  C. Anastasiou, D. Kosower, and Z. Kunszt for helpful discussions.
RB thanks the Galileo Galilei Institute for Theoretical Physics for
hospitality and the INFN for partial support during the workshop
``Advancing Collider Physics.''
 She is supported by Stichting FOM. BF would like to thank the
 Imperial College, London where this project started. He is
 supported by Qiu-Shi Professor Fellowship from Zhejiang University,
 China.

\appendix

\section{\label{spinorintegration}Setting up the cut integral}

In this appendix we briefly review the first steps in spinor integration, in the context of $d$-dimensional unitarity,  leading from
equation (\ref{I-inte}) to equation (\ref{cut-refined}).  For a fuller
discussion of this technique, see \cite{Anastasiou:2006gt}.
Within the four-dimensional helicity scheme, we apply (\ref{fdh}) and (\ref{def-u}).  In the integrand, $p$ is replaced by $\W\ell$, and the measure is transformed as follows:
\bean \int {d^{4-2\eps} p\over (2\pi)^{4-2\eps}} & = & \int {d^{4} \W\ell\over (2\pi)^4}
{(4\pi)^{\eps}\over \Gamma(-\eps)} \int  d\mu^2 ~
(\mu^2)^{-1-\eps},\eean
We now drop the factor $ {(4\pi)^\eps \over (2\pi)^4\Gamma(-\eps)}
\left({K^2 \over 4}\right)^{-\eps}$, which is universal and common to cuts of
amplitudes and master integrals.
Following \cite{Anastasiou:2006jv, Anastasiou:2006gt}, we further decompose the 4-dimensional momentum into a null component and a component proportional to the cut momentum $K$.
\bean \W \ell= \ell+z K,~~~~~\ell^2=0,~~~\Longrightarrow \int d^4\W
\ell= \int dz ~d^4\ell ~\delta^+(\ell^2) (2 \ell \cdot K).
\eean
While changing the variable $\mu$ to $u$ with (\ref{def-u}), we note
that the kinematics of the unitarity cut constrain the integration domain to be $u\in [0,1]$.
Our cut integral (\ref{I-inte}) can now been rewritten as the following
expression.
\bean
& & \int_0^1 du~ u^{-1-\eps}
\int dz~ (1-2z)  \delta(z(1-z)- {u \over 4})
\\ & & \int d^4\ell~
\delta^+(\ell^2)\delta( (1-2z)K^2 -2 \ell\cdot K) \frac{
\prod_{i=1}^{M} (-2 P_i \cdot (\ell+zK) )
}{\prod_{j=1}^N (K_j^2-z
(2K_j\cdot
K)-2\ell\cdot K_j)}
\eean
Notice that the $z$-integral can now be done with the first delta function.
In fact, the kinematics of the unitarity cut require us to choose
exactly one solution for $z$.  If we take $K>0$, then
\bea z={1- \sqrt{1-u}\over
2},~~~~{\rm or~equivalently,}~~~~1-2z=\sqrt{1-u}.~~~~\Label{z-u} \eea
Now we change to spinor variables with \cite{Cachazo:2004kj}
\bea
\ell = t\lambda\W\la,~~~\Label{spinormeasure}
\eea
where $t$ takes nonnegative real values, and $\lambda$ and $\W\la$ are homogeneous spinors.  The measure transforms as
\bean
\int d^4\ell~ \delta^{(+)}(\ell^2) ~ (\bullet ) =
\int_0^{\infty}dt~t\int_{\tilde\lambda=\bar\lambda}\vev{\lambda~
d\lambda}[\tilde\lambda~d\tilde\lambda] ( \bullet ).
\eean
The domain of integration of $t$ is again consistent with the
kinematic region of the unitarity cut.  From here on, we use
$\ket{\ell}$ and $|\ell]$ interchangeably with $\lambda$ and
$\tilde\lambda$.  We have now arrived at the following expression:
\bean
C =
\int du~ u^{-1-\eps}  \int \vev{\ell~d\ell}[\ell~d\ell]\int t~
dt~ \delta( (1-2z)K^2 +t \gb{\ell|K|\ell}) \frac{ \prod_{i=1}^{M}
(-z (2 K \cdot P_i)
+t \gb{\ell|P_i|\ell})}{\prod_{j=1}^N (K_j^2-z
(2K_j\cdot K)+t \gb{\ell|K_j|\ell})}
\eean
Finally, we use the remaining delta function to perform the integral
over the variable $t$. With the substitution (\ref{z-u}), the result is equation (\ref{cut-refined}).

\section{\label{derivation}Derivation of the formulas for coefficients}

In this appendix we outline the derivation of the main results of this
paper, which are the formulas (\ref{box-exp}), (\ref{tri-exp}) and (\ref{bub-exp}).  Our technique is the type of spinor integration carried out in
\cite{Britto:2005ha,Britto:2006sj,Britto:2006fc}, but we stress that understanding these techniques is
unnecessary for applying the results.  Indeed, equivalent formulas
have already appeared in \cite{Britto:2006fc}.\footnote{The formulas
  for coefficients given in \cite{Britto:2006fc} differ from the ones
  given here by a factor of $\sqrt{1-u}$.  This comes from the
  convention of our starting point (\ref{I-inte}) or equivalently
  (\ref{cut-refined}), where this factor appears explicitly.}  The difference is that
our starting point is now the raw unitarity cut integral, before
converting the loop momentum to spinor variables.  In the final
formulas, we now explicitly evaluate the residue at multiple poles.  Additionally, the present versions of the formulas
feature substantial simplification of the bubble coefficients.

Our foundation here is the framework laid out in \cite{Britto:2006fc}
and its references.  Let us briefly recall the key ideas.
The general integrand given as the starting point in
\cite{Britto:2006fc} is \footnote{We have redefined the index $n$ for consistency.}
\bea I_{term}={ G(\la) \prod_{j=1}^{n+k} [a_j~\ell]\over
\gb{\ell|K|\ell}^{n+2}
\prod_{p=1}^{k}\gb{\ell|Q_p|\ell}}.~~\Label{gen-form}\eea
Comparing with the expression (\ref{cut-refined}), we see that we will
take $G(\la)$ to be constant and $[a_j|=\langle \ell | R_j|.$  The
idea of spinor integration is to rewrite the integral so that we can
carry it out with the residue theorem.  The next step, therefore, is
to isolate poles by splitting the denominator factors with spinor
identities such as
\bea {[a~\ell]\over \gb{\ell|Q_1|\ell}\gb{\ell|Q_2|\ell}}
={\tgb{a|Q_1|\ell}\over \vev{\ell|Q_2
Q_1|\ell}\gb{\ell|Q_1|\ell}}+{\tgb{a|Q_2|\ell}\over \vev{\ell|Q_1
Q_2|\ell}\gb{\ell|Q_2|\ell}}.~~\Label{Ge-split}
\eea
This procedure is applied to the amplitude on one hand and the master
integrals on the other.  By matching functional forms, we extract the
coefficients.

\subsection{Box}

The formula (\ref{box-exp}) for a box coefficient is trivially related to the one given in
\cite{Britto:2006fc}.  We only need to observe that now that we take
$G(\la)$ to be constant and $[a_j|=\langle \ell | R_j|$ in
(\ref{gen-form}), the poles from the factors $\vev{\ell|Q_sQ_r|\ell}$
are inserted into $[a_j|$ as well when we evaluate the residue.

\subsection{\label{tri-der}Triangle}

For a triangle associated to momenta $K_s,K$, the coefficient was
found in \cite{Britto:2006fc} to be the difference of the two residues
from the poles in
 $\vev{\ell|Q_s
K|\ell}^{n+2}$ of the following function:\footnote{Here we have again
redefined $n$ and made substitutions for $G(\la)$ and $[a_j|$.}
\bea  {(-1)^n
(K^2)^{1+n}\sqrt{\Delta_s}\over 2}{\prod_{j=1}^{k+n} \vev{\ell|R_j
Q_s|\ell}\over \vev{\ell|Q_s K|\ell}^{n+2} \prod_{t=1,t\neq s}^k \vev{\ell|Q_t Q_s
|\ell}}~.~~\Label{Tri-exp}
\eea

The quantities $\Delta_s, P_{s,1},P_{s,2}$ were defined in
(\ref{tri-null}) specifically to deal with the factor  $\vev{\ell|Q_s
K|\ell}$ by identifying the poles explicitly.
With those definitions, we find an identity that separates the two poles:
\bea
\vev{\ell|Q_s K|\ell}={\vev{\ell~P_{s,1}}\vev{\ell~P_{s,2}}}{K^2 [P_{s,1}~P_{s,2}]\over\sqrt{\Delta_s}}~~~~\Label{QK-factor}.
\eea

Now consider the residue from a multiple pole, in an expression of the
form
\bean
{1\over
\vev{\ell~\eta}^n}{N(\ket{\ell},|\ell])\over D(\ket{\ell},|\ell])}.
\eean
We can start by substituting $|\ell]=|\eta]$, so we are dealing with
the holomorphic function
\bean
{1\over
\vev{\ell~\eta}^n}{N(\ket{\ell},|\eta])\over D(\ket{\ell},|\eta])}.
\eean

For an arbitrary auxiliary spinor $\zeta$, we have the following identity.
\bea \frac{1}{\vev{\ell~(\eta-\tau
\zeta)}^n}=\frac{d^{n-1}}{d\tau^{n-1}}\left.\left(\frac{1}{(n-1)!\vev{\ell~\zeta}^{n-1}}
\frac{1}{\vev{\ell~(\eta-\tau \zeta)}}\right)\right|_{\tau\to 0}\eea
Thus we find
\bea \frac{1}{\vev{\ell~(\eta-\tau
\zeta)}^n}=\frac{d^{n-1}}{d\tau^{n-1}}\left.\left(\frac{1}{(n-1)!\vev{\ell~\zeta}^{n-1}}
\frac{1}{\vev{\ell~(\eta-\tau \zeta)}}{N(\ket{\ell},|\eta])\over
D(\ket{\ell},|\eta])}\right)\right|_{\tau\to 0}~.\eea
Now we extract the  residue at the  single pole
$\vev{\ell~(\eta-\tau \zeta)}$ before taking the derivative.  We find
\bea
\frac{d^{n-1}}{d\tau^{n-1}}\left.\left(\frac{1}{(n-1)!\vev{\eta~\zeta}^{n-1}}
{N(\ket{\eta-\tau \zeta},|\eta])\over D(\ket{\eta-\tau
\zeta},|\eta])}\right)\right|_{\tau\to 0}~.~~~\Label{multi-pole}\eea

To obtain the residues from the factor $\vev{\ell|Q K|\ell}^n$, we use
equation (\ref{QK-factor}) to rewrite it in terms of two multiple
poles.\footnote{Throughout the rest of this derivation, we drop the
  subscript $s$ to avoid cluttering the formulas.} Then
we apply (\ref{multi-pole})
to compute the two residues as follows:
\bean  R_1 & = & \frac{d^{n-1}}{d\tau_1^{n-1}}\left.\left( \frac{1}{(n-1)!
\vev{P_1~\zeta_1}^{n-1}}\frac{(K^2)^n N(\ket{P_1-\tau_1
\zeta_1},|P_1])}{(\sqrt{\Delta})^n
[P_1~P_2]^n\vev{P_1-\tau_1 \zeta_1, P_2}^n D(\ket{P_1-\tau_1 \zeta_1},|P_1])}\right)\right|_{\tau\to 0}, \\
R_2 & = & \frac{d^{n-1}}{d\tau_2^{n-1}}\left.\left( \frac{1}{(n-1)!
\vev{P_2~ \zeta_2}^{n-1}}\frac{(K^2)^n N(\ket{P_2-\tau_2
\zeta_2},|P_2])}{(\sqrt{\Delta})^n [P_1~P_2]^n\vev{P_2-\tau_2
\zeta_2, P_1}^n D(\ket{P_2-\tau_2
\zeta_2},|P_2])}\right)\right|_{\tau\to 0}.
\eean
where we can choose different auxiliary spinors $\zeta_1,\zeta_2$ for
the two poles. To simplify
further, we choose $\ket{\zeta_1}=\ket{P_2}$ and $\ket{\zeta_2}=\ket{P_1}$  and use the identity
$[P_1~P_2]\vev{P_1~P_2}=
-\Delta/K^2$.  Finally we find
\bea  R_1 & = & \frac{(-1)^n}{(\sqrt{\Delta})^n}\frac{1}{(n-1)!
\vev{P_1~P_2}^{n-1}}\frac{d^{n-1}}{d\tau_1^{n-1}}\left.\left( \frac{
N(\ket{P_1-\tau_1 P_2},|P_1])}{ D(\ket{P_1-\tau_1 P_2},|P_1])}
\right)\right|_{\tau\to 0},~~~\Label{multi-R1} \\
R_2 & = & -\frac{(-1)^n}{(\sqrt{\Delta})^n}\frac{1}{(n-1)!
\vev{P_1~P_2}^{n-1}}\frac{d^{n-1}}{d\tau_2^{n-1}}\left.\left( \frac{
N(\ket{P_2-\tau_2 P_1},|P_2])}{ D(\ket{P_2-\tau_2
P_1},|P_2])}\right)\right|_{\tau\to 0}.~~~\Label{multi-R2}\eea
Using  (\ref{multi-R1}) and (\ref{multi-R2}) with our original expression (\ref{Tri-exp}), we get the
formula (\ref{tri-exp}) for the triangle coefficient.
That formula may look as though it is not completely explicit
because we still need to perform a differentiation.  But this is
easily done in a symbolic manipulation program.
We do offer explicit formulas in Appendix \ref{closed-triangle} but do
not expect those to be more useful.

Recall that the final result must be
a rational function, so the square roots from $\sqrt{\Delta_s}$ should
eventually combine into polynomial expressions.

\subsection{Bubble}

Our derivation here parallels the one in \cite{Britto:2006fc}, but the splitting
identities are more systematic and the final formula is now written explicitly.

First, we would like to split the denominator factors in
(\ref{gen-form}) using the following
generalization of (\ref{Ge-split}):

\bea  {\prod_{j=1}^{k-1}[a_j~\ell]\over \prod_{i=1}^k
\gb{\ell|Q_i|\ell}}=\sum_{i=1}^k {1\over
\gb{\ell|Q_i|\ell}}{\prod_{j=1}^{k-1} \tgb{a_j|Q_i|\ell}\over
\prod_{j=1,j\neq i}^k \vev{\ell|Q_j
Q_i|\ell}}~~\Label{Split-ob-1}\eea
This formula is applicable when and only when all $Q_i$ and $K$ are
different. To use it, we deform (\ref{gen-form}) by introducing
small independent parameters  $s_i$, $i=1,...,n+1$ and a {\sl real}
null vector $\eta$.
\bea { G(\la) \prod_{j=1}^{n+k} [a_j~\ell]\over
\gb{\ell|K|\ell}\prod_{i=1}^{n+1}\gb{\ell|K+ s_i \eta|\ell}
\prod_{p=1}^{k}\gb{\ell|Q_p|\ell}}~~~\Label{def-form}\eea
The final result will be recovered by taking the limit
 $s_i\to 0$.

Now we can apply (\ref{Split-ob-1}) to
(\ref{def-form}) to find the following expression:
\bea & & \sum_{i=1}^{n+1} {1\over \gb{\ell|K|\ell}\gb{\ell|K+ s_i
\eta|\ell}} {G(\la)\prod_{j=1}^{n+k}\tgb{a_j| K+ s_i \eta|\ell}\over
\prod_{q=1,q\neq i}^{n+1} \vev{\ell| (K+ s_q \eta)( K+ s_i \eta)
|\ell}\prod_{p=1}^k \vev{\ell|Q_p ( K+ s_i \eta)|\ell}}~~~\Label{lineone}
\\ & & + \sum_{i=1}^{k} {1\over \gb{\ell|K|\ell}\gb{\ell|Q_i|\ell}}
{(G(\la)\prod_{j=1}^{n+k}\tgb{a_j| Q_i|\ell}\over \prod_{q=1}^{n+1}
\vev{\ell| (K+ s_q \eta)Q_i |\ell}\prod_{r=1,r\neq i}^k
\vev{\ell|Q_{r} Q_i|\ell}}~~~\Label{linetwo} \eea
We can see that the $s_i\to 0$ limit is smooth in the second line,
resulting in terms of the form  $ F_i(\la)/(
\gb{\ell|K|\ell}\gb{\ell|Q_i|\ell})$.  We know from \cite{Britto:2006sj} that these
terms yield pure logarithms, in this case for the triangle integrals
associated with momenta $K$ and $K_i$.  So we restrict our attention
to the first line, (\ref{lineone}).  Rewrite it as
\bean \sum_{i=1}^{n+1} {1\over \gb{\ell|K|\ell}\gb{\ell|K+ s_i
\eta|\ell}} {G(\la)\prod_{j=1}^{n+k}\tgb{a_j| K+ s_i \eta|\ell}\over
\vev{\ell|K \eta|\ell}^{n}\prod_{q=1,q\neq i}^{n+1}
(s_i-s_q)\prod_{p=1}^k \vev{\ell|Q_p ( K+ s_i \eta)|\ell}}\eean
Now  we can must take  the limit  $s_i\to 0$ carefully.
We find that the bubble coefficient is
\bea \sum_{q=0}^n \left. {(-1)^q\over q!} {d^q
B_{n,n-q}(s)\over ds^q}\right|_{s=0},~~~~\Label{gen-n}\eea
where we have defined the function
\bea B_{n,t}(s) \equiv {\gb{\ell|\eta|\ell}^t\over
\gb{\ell|K|\ell}^{2+t}}{ G(\la)\prod_{j=1}^{n+k}
\tgb{a_j|K+s\eta|\ell}\over \vev{\ell|\eta K|\ell}^n \prod_{p=1}^k
\vev{\ell| Q_p(K+s\eta)|\ell}}~.~~~\Label{Bnt}\eea
The fact that (\ref{gen-n}) represents the bubble coefficient can be
proved by induction.  The case $n=0$ is trivial.  Assume that it is true
for $n$, and let us now introduce a single parameter $\W s$ to rewrite
(\ref{def-form}) as
\bea { G(\la) \prod_{j=1}^{n+k+1} [a_j~\ell]\over
\gb{\ell|K|\ell}^{n+2} \gb{\ell|K+\W s\eta|\ell}
\prod_{p=1}^{k}\gb{\ell|Q_p|\ell}}.
\eea
We now treat the factor $\gb{\ell|K+\W s\eta|\ell}$ on the same
footing as $\gb{\ell|Q_p|\ell}$ and apply the result for $n$.
The bubble contribution can be expressed as a sum of two terms $I_1$
and $I_2$.  The first term is
\bean I_1 & = & { 1\over \gb{\ell|K|\ell}\gb{\ell|K+\W s\eta|\ell}}
{G(\la) \prod_{j=1}^{n+k+1} \tgb{a_j|K+\W s \eta|\ell}\over
\vev{\ell|K (K+\W s\eta)|\ell}^{n+1} \prod_{p=1}^k \vev{\ell|Q_p
(K+\W s\eta)|\ell}} \\
& = & {(-1)^{n+1} \over \W s^{n+1}} \left( \sum_t  (-1)^t \W s^t
{\gb{\ell|\eta|\ell}^t \over \gb{\ell|K|\ell}^{t+2}}\right) {G(\la)
\prod_{j=1}^{n+k+1} \tgb{a_j|K+\W s \eta|\ell}\over \vev{\ell|\eta K
|\ell}^{n+1} \prod_{p=1}^k \vev{\ell|Q_p (K+\W s\eta)|\ell}}\eean
After taking the $\W s\to 0$ limit, we have
\bea I_1 & = & \sum_{a=0}^{n+1} \sum_{t=0}^{n+1-a}  {(-1)^{n+1+t}
\over \W s^{n+1-t-a}a!} {d^a { B}_{n+1,t}(\W s=0)\over d \W s^a},\eea
where $B_{n,t}(s)$ is defined by (\ref{Bnt}).

The second contribution is
\bean I_2 & = & \sum_{q=0}^n {(-1)^q\over q!} {d^q \W
B_{n,n-q}(s=0)\over ds^q},\eean
where
\bea \W B_{n,t}(s) & \equiv & {\gb{\ell|\eta|\ell}^t\over
\gb{\ell|K|\ell}^{2+t}}{ G(\la)\prod_{j=1}^{n+k+1}
\tgb{a_j|K+s\eta|\ell}\over \vev{\ell|\eta K|\ell}^n \vev{\ell|(K+\W
s \eta) (K+s \eta)|\ell}\prod_{p=1}^k \vev{\ell|
Q_p(K+s\eta)|\ell}}~\nonumber \\& = &
 {1\over (\W s-s)} { B}_{n+1,t}(s). \eea
We must take the $s\to 0$ limit before $\W s\to 0$, so first we substitute
\bean \left.{d^q \W B_{n,n-q}(s)\over ds^q}\right|_{s=0}= \sum_{b=0}^q{q!\over
(q-b)!\W s^{1+b}} {d^{q-b} \over ds^{q-b}}{ B}_{n+1,t}(s=0)\eean
to find
\bean I_2 & = & \sum_{q=0}^n \sum_{b=0}^q{(-1)^q \over (q-b)!\W s^{1+b}}
{d^{q-b}
\over ds^{q-b}}{ B}_{n+1,n-q}(s=0),
\eean
or equivalently,
\bea I_2 & = & \sum_{a=0}^n \sum_{t=0}^{n-a} {(-1)^{n-t} \over a!\W
s^{1+n-t-a}} {d^{a} \over ds^{a}}{ B}_{n+1,t}(s=0)\eea
Now it is easy to see that $I_1+I_2$ is nonzero only if
 $a+t=n+1$.  Therefore we can write
\bea I_1+ I_2= \sum_{a=0}^{n+1}  {(-1)^{a} \over a!} {d^{a} \over
ds^{a}}{ B}_{n+1,n+1-a}(s=0),\eea
and thus we have proved the formula (\ref{gen-n}) for $n+1$.

Now that we have established that the bubble coefficient comes from
(\ref{gen-n}) with the definition (\ref{Bnt}), we need to identify the
poles and find the residues.

Rewrite the integrand  (\ref{gen-n}) as a total derivative by using
\bean
[\ell~d\ell] B_{n,t}(s) =
[d\ell~\partial_\ell] \left( {G(\lambda)\over (t+1)} {\gb{\ell|\eta|\ell}^{t+1}\over
\gb{\ell|K|\ell}^{t+1}}{\prod_{j=1}^{n+k}
\tgb{a_j|K+s\eta|\ell}\over \vev{\ell|\eta K|\ell}^{n+1}
\prod_{p=1}^k \vev{\ell| Q_p(K+s\eta)|\ell}} \right) \eean
Now let us specialize to the integrand of (\ref{cut-refined}), so
that $G(\la)$ is constant and $[a_j|=\langle \ell | R_j|.$ Now we
define \footnote{The $B_{n,t}(s)$ is the splitting result while
${\cal B}_{n,t}(s)$ is after writing into total derivative,
i.e.,$B_{n,t}(s)=[d\W \la~\partial_{\W \la}]{\cal B}_{n,t}(s)$.}
\bea {\cal B}_{n,t}(s) \equiv {1\over (t+1)}
{\gb{\ell|\eta|\ell}^{t+1}\over
\gb{\ell|K|\ell}^{t+1}}{\prod_{j=1}^{n+k} \vev{\ell|R_j
(K+s\eta)|\ell}\over \vev{\ell|\eta K|\ell}^{n+1} \prod_{p=1}^k
\vev{\ell| Q_p(K+s\eta)|\ell}}~~~\Label{cal-Bnt} \eea
 Here it is important that $\eta$ be completely generic,
so that there are no accidental degeneracies.  For an alternative
approach, see Subsection \ref{specialcase}.  There are three kinds of
poles.  The first, at $\ell=\eta$, has no residue because the
numerator factor $\gb{\ell|\eta|\ell}^{t+1}$ becomes zero.
The second, at $\ket{\ell}=K|\eta]$, is a multiple pole of the type
discussed in \ref{tri-der}, so we see that its residue is (\ref{cal-B-0}).
The last kind of pole is from the factor $\vev{\ell|Q_r (K+s \eta)|\ell}$.
Here we perform a series expansion in the parameter $s$, which we will
ultimately set to zero.  The expansion is
\bea {1\over \vev{\ell|Q_r (K+s\eta)|\ell}} = \sum_{a=0} (-s)^a
{\vev{\ell|Q_r \eta|\ell}^a\over \vev{\ell|Q_r K|\ell}^{a+1}}~. \eea
The residue is then ${\cal B}_{n,t}^{(r;a;1)}(s)-{\cal B}_{n,t}^{(r;a;2)}(s)$, with the
definitions given in (\ref{cal-B-r-1}) and (\ref{cal-B-r-2}).

Combining these contributions, we find that the sum of the residues at
poles of ${\cal
B}_{n,t}(s)$ is
\bea {\cal B}_{n,t}^{(0)}(s)+\sum_{r=1}^k
\sum_{a=0} (-s)^a \left({\cal B}_{n,t}^{(r;a;1)}(s)-{\cal
B}_{n,t}^{(r;a;2)}(s)\right).~~~\Label{RB}\eea

Feeding (\ref{RB}) into (\ref{gen-n}) and simplifying the
result
gives us our final expression for the
bubble coefficient, (\ref{bub-exp}).

\subsubsection{\label{specialcase}A special choice of $\eta$}

In this appendix, we describe the consequences of choosing $\eta=K_1$ in the case where $K_1^2$.  This choice may be convenient for small examples worked by hand, but we emphatically recommend choosing a generic $\eta$ wherever possible.

The reason that such a special choice of $\eta$ presents a problem is the following.  From (\ref{RQ}), we can see that
\bean \vev{\ell|Q_1 K|\ell}=-(1-2z)\vev{\ell|K_1K|\ell}=-(1-2z)\vev{\ell|\eta K|\ell}
\eean
Therefore, in the expression  (\ref{cal-Bnt}), the poles from
 $\vev{\ell|\eta K|\ell}$ and  $\vev{\ell|Q_1
(K+s\eta)|\ell}$, will overlap, and the way we read off
their residues should be modified respectively.

In this special case, it is easy to see that
\bean {1\over \vev{\ell|Q_1 (K+s\eta)|\ell}}& = & \sum_{a=0} (-s)^a
{\vev{\ell|Q_1 \eta|\ell}^a\over \vev{\ell|Q_1 K|\ell}^{a+1}}=
- {1\over \vev{\ell|\eta
K|\ell}} { 1\over (1-2z) - s z {(2K\cdot \eta)\over K^2}} \eean
Now instead of  (\ref{cal-Bnt}), we have
\bea {\cal B}_{n,t}(s) \equiv -{ 1\over (1-2z) - s z {(2K\cdot
\eta)\over K^2}}{1\over (t+1)} {\gb{\ell|\eta|\ell}^{t+1}\over
\gb{\ell|K|\ell}^{t+1}}{\prod_{j=1}^{n+k} \vev{\ell|R_j
(K+s\eta)|\ell}\over \vev{\ell|\eta K|\ell}^{n+2} \prod_{p=2}^k
\vev{\ell| Q_p(K+s\eta)|\ell}}.~~~\Label{spe-cal-Bnt} \eea
Continuing this way, we find
\bea {\cal B}_{n,t}^{(0)}(s) &\equiv& -{d^{n+1}\over d\tau^{n+1}}\left(
{ 1\over (1-2z) - s z {(2K\cdot \eta)\over K^2}}{[\eta|\W \eta
K|\eta]^{-n-1}\over (t+1) (n+1)!} \left( {(2\eta\cdot K)\over
K^2}\right)^{t+1}\right. \nonumber
\\ & & \left.\left.
{\prod_{j=1}^{n+k} \vev{\ell|R_j
(K+s\eta)|\ell}\over \vev{\ell~\eta}^{n+2} \prod_{p=2}^k \vev{\ell|
Q_p(K+s\eta)|\ell}}|_{\ket{\ell}\to |K-\tau \W \eta|\eta]
}\right)\right|_{\tau\to 0},~~~\Label{spe-cal-B-0}\eea
\bea {\cal B}_{n,t}^{(r;a;1)}(s) & \equiv & { 1\over (1-2z) - s z
{(2K\cdot \eta)\over K^2}}{(-1)^{a}\over \sqrt{\Delta_r}^{a+1} a!
\vev{P_{r,1}~P_{r,2}}^a}{d^a \over d\tau^a} \left({1\over (t+1)}
{\gb{\ell|\eta|\ell}^{t+1}\over \gb{\ell|K|\ell}^{t+1}}\right.
\nonumber \\ & & \times \left.\left. {\vev{\ell|Q_r
\eta|\ell}^a\prod_{j=1}^{n+k} \vev{\ell|R_j (K+s\eta)|\ell}\over
\vev{\ell|\eta K|\ell}^{n+2} \prod_{p=1,p\neq r}^k \vev{\ell|
Q_p(K+s\eta)|\ell}}\right)\right|_{|\ell]=|P_{r,1}],\ket{\ell}=\ket{P_{r,1}}-\tau
\ket{P_{r,2}}}~~~\Label{spe-cal-B-r-1}\eea
\bea {\cal B}_{n,t}^{(r;a;2)}(s) & \equiv & { 1\over (1-2z) - s z
{(2K\cdot \eta)\over K^2}}{(-1)^{a}\over \sqrt{\Delta_r}^{a+1} a!
\vev{P_{r,1}~P_{r,2}}^a}{d^a \over d\tau^a} \left({1\over (t+1)}
{\gb{\ell|\eta|\ell}^{t+1}\over \gb{\ell|K|\ell}^{t+1}}\right.
\nonumber \\ & & \times \left.\left. {\vev{\ell|Q_r
\eta|\ell}^a\prod_{j=1}^{n+k} \vev{\ell|R_j (K+s\eta)|\ell}\over
\vev{\ell|\eta K|\ell}^{n+2} \prod_{p=1,p\neq r}^k \vev{\ell|
Q_p(K+s\eta)|\ell}}\right)\right|_{|\ell]=|P_{r,2}],\ket{\ell}=\ket{P_{r,2}}-\tau
\ket{P_{r,1}}}~~~\Label{spe-cal-B-r-2}\eea
and the coefficient is given by
\bea C[K]_n = (K^2)^{1+n} \sum_{q=0}^n {(-1)^q\over q!} {d^q \over
ds^q}\left.\left( {\cal B}_{n,n-q}^{(0)}(s)+\sum_{r=2}^k\sum_{a=q}^n
\left({\cal B}_{n,n-a}^{(r;a-q;1)}(s)-{\cal
B}_{n,n-a}^{(r;a-q;2)}(s)\right)\right)\right|_{s=0}.~~~~\Label{spe-Re-gen-n-1}\eea
where the definitions of the functions ${\cal B}$ must be taken from (\ref{spe-cal-B-0}),
(\ref{spe-cal-B-r-1}) and (\ref{spe-cal-B-r-2}).

\section{\label{closed-triangle}Closed forms for triangle coefficients}

We have given the general expression for coefficients of triangles
as a formula involving a multiple derivative (\ref{tri-exp}). We can also carry out the differentiation explicitly. For practical purposes, we need to consider only cases with  $n\leq 2$.

When $n\leq -2$, the contribution is simply zero.

When $n=-1$, there is no derivative, so the result is just
\bea C[Q_s,K]_{n=-1} & = & {1\over 2}\left( {\prod_{j=1}^{k-1}
\gb{P_{s,1}|R_j|P_{s,2}}\over \prod_{t=1,t\neq s}^k \gb{P_{s,1}|Q_t
|P_{s,2}}} \right)~~~~~\Label{tri-exp-n=-1}\eea

When $n=0$ it is given by
\bea C[Q_s,K]_{n=0} & = & {K^2\over 2
\Delta_s}\left\{{\prod_{j=1}^{k} \gb{P_{s,1}|R_j|P_{s,2}}\over
\prod_{t=1,t\neq s}^k \gb{P_{s,1}|Q_t |P_{s,2}}} \left( \sum_{j=1}^k
{(2 Q_s\cdot K)(2R_j \cdot Q_s)-2 Q_s^2 (2R_j\cdot K)\over
\gb{P_{s,1}|R_j|P_{s,2}}} \right.\right. \nonumber \\ & & \left.
\left.-\sum_{t=1,t\neq s}^k {(2 Q_s\cdot K)(2Q_t \cdot Q_s)-2 Q_s^2
(2Q_t\cdot K)\over \gb{P_{s,1}|Q_t|P_{s,2}}}\right)+
\{P_{s,1}\leftrightarrow
P_{s,2}\}\right\}~~~~~\Label{tri-exp-n=0}\eea

When $n=1$ it will be
\bea C[Q_s,K]_{n=1} & = & {(K^2)^2\over 4
\Delta_s^2}\left\{{\prod_{j=1}^{k+1} \gb{P_{s,1}|R_j|P_{s,2}}\over
\prod_{t=1,t\neq s}^k \gb{P_{s,1}|Q_t |P_{s,2}}} \left[\left(
\sum_{j=1}^{k+1} {(2 Q_s\cdot K)(2R_j \cdot Q_s)-2 Q_s^2 (2R_j\cdot
K)\over \gb{P_{s,1}|R_j|P_{s,2}}} \right.\right. \right.\nonumber \\
& &  \left.-\sum_{t=1,t\neq s}^k {(2 Q_s\cdot K)(2Q_t \cdot Q_s)-2
Q_s^2 (2Q_t\cdot K)\over \gb{P_{s,1}|Q_t|P_{s,2}}}\right)^2 \nonumber \\
& & +\sum_{j=1}^{k+1} {-[(2 Q_s\cdot K)(2R_j \cdot Q_s)-2 Q_s^2
(2R_j\cdot K)]^2+2 Q_s^2 K^2
\gb{P_{s,1}|R_j|P_{s,2}}\gb{P_{s,2}|R_j|P_{s,1}}\over
\gb{P_{s,1}|R_j|P_{s,2}}^2} \nonumber \\& & - \sum_{t=1,t\neq s}^k
{-[(2 Q_s\cdot K)(2Q_t \cdot Q_s)-2 Q_s^2 (2Q_t\cdot K)]^2+2 Q_s^2
K^2 \gb{P_{s,1}|Q_t|P_{s,2}}\gb{P_{s,2}|Q_t|P_{s,1}}\over
\gb{P_{s,1}|Q_t|P_{s,2}}^2} \nonumber \\& & \left.+
\{P_{s,1}\leftrightarrow P_{s,2}\}\right\}
~~~~~\Label{tri-exp-n=1}\eea

For $n=2$ the result is
\bea C[Q_s,K]_{n=2} & = & {(K^2)^3\over 12
\Delta_s^3}\left\{{\prod_{j=1}^{k+2} \gb{P_{s,1}|R_j|P_{s,2}}\over
\prod_{t=1,t\neq s}^k \gb{P_{s,1}|Q_t |P_{s,2}}} ({\cal A}^3+ 3
{\cal A}{\cal B}+{\cal C}) + \{P_{s,1}\leftrightarrow
P_{s,2}\}\right\},~~~~~\Label{tri-exp-n=2}\eea
 where we have
defined
\bean {\cal A} & = & \sum_{j=1}^{k+2} {(2 Q_s\cdot K)(2R_j \cdot
Q_s)-2 Q_s^2 (2R_j\cdot K)\over \gb{P_{s,1}|R_j|P_{s,2}}}
-\sum_{t=1,t\neq s}^k {(2 Q_s\cdot K)(2Q_t \cdot Q_s)-2 Q_s^2
(2Q_t\cdot K)\over
\gb{P_{s,1}|Q_t|P_{s,2}}} \\
{\cal B} & = & - \sum_{j=1}^{k+2} {[(2 Q_s\cdot K)(2R_j \cdot Q_s)-2
Q_s^2 (2R_j\cdot K)]^2+2 Q_s^2 K^2
\gb{P_{s,1}|R_j|P_{s,2}}\gb{P_{s,2}|R_j|P_{s,1}}\over
\gb{P_{s,1}|R_j|P_{s,2}}^2}\nonumber \\ & & + \sum_{t=1,t\neq s}^k
{[(2 Q_s\cdot K)(2Q_t \cdot Q_s)-2 Q_s^2 (2Q_t\cdot K)]^2 - 2 Q_s^2
K^2 \gb{P_{s,1}|Q_t|P_{s,2}}\gb{P_{s,2}|Q_t|P_{s,1}}\over
\gb{P_{s,1}|Q_t|P_{s,2}}^2}\\
{\cal C} & = & \sum_{j=1}^{k+2} {[(2 Q_s\cdot K)(2R_j \cdot Q_s)-2
Q_s^2 (2R_j\cdot K)]^3-3 Q_s^2 K^2
\gb{P_{s,1}|R_j|P_{s,2}}\gb{P_{s,2}|R_j|P_{s,1}}\over
\gb{P_{s,1}|R_j|P_{s,2}}^2} \\
& & {2[(2 Q_s\cdot K)(2R_j \cdot Q_s)-2 Q_s^2 (2R_j\cdot K)]\over
\gb{P_{s,1}|R_j|P_{s,2}}}-  \sum_{t=1,t\neq s}^k {2[(2 Q_s\cdot
K)(2Q_t \cdot Q_s)-2 Q_s^2 (2Q_t\cdot K)]\over
\gb{P_{s,1}|Q_t|P_{s,2}}} \\
& &  {[(2 Q_s\cdot K)(2Q_t \cdot Q_s)-2 Q_s^2 (2Q_t\cdot K)]^2-3
Q_s^2 K^2 \gb{P_{s,1}|Q_t|P_{s,2}}\gb{P_{s,2}|Q_t|P_{s,1}}\over
\gb{P_{s,1}|Q_t|P_{s,2}}^2}\eean
%


\end{document}